\documentclass[%
 reprint,
 amsmath,amssymb,
 aps,
 superscriptaddress,altaffilletter
]{revtex4-2}
\usepackage{graphicx}% Include figure files
\usepackage{dcolumn}% Align table columns on decimal point
\usepackage{bm}% bold math
\usepackage{ulem}
\usepackage{siunitx}
\usepackage{xcolor}

\makeatletter

\begin{document}

%\preprint{APS/123-QED}

%%%%%%%%%%%%%%% TITLE

\title{Stability of a tilted granular monolayer: \\How many spheres can we pick before the collapse?}% Force line breaks with \\

%%%%%%%%%%%%%%% Afiliaciones

\author{Eduardo Rojas}
\email{eduardo.rojas@uantof.cl}
\affiliation{Department of Mechanical Engineering, Universidad de Antofagasta. Antofagasta, Chile.}

\author{H\'ector Alarc\'on}
\affiliation{Departamento de F\'isica y Qu\'imica, Facultad de Ingenier\'ia, Universidad Aut\'onoma de Chile. Santiago, Chile.}
%\affiliation{Instituto de Ciencias de la Ingenier\'ia, Universidad de O'Higgins. Rancagua, Chile.}

\author{Vicente Salinas}
\affiliation{Instituto de Ciencias Qu\'imicas Aplicadas, Facultad de Ingenier\'ia, Universidad Aut\'onoma de Chile. Santiago, Chile.}

\author{Gustavo Castillo}
\affiliation{Instituto de Ciencias de la Ingenier\'ia, Universidad de O'Higgins. Rancagua, Chile.}

\author{Pablo Guti\'errez}
\email{pablo.gutierrez@uoh.cl}
\affiliation{Instituto de Ciencias de la Ingenier\'ia, Universidad de O'Higgins. % Av. Libertador Bernardo O'Higgins 611, 
Rancagua, Chile.}

\date{\today}% It is always \today, today,

%%%%%%%%%%%%%%% ABSTRACT

\begin{abstract}

The triggering of avalanches is investigated using discrete element simulations for a process of random extraction of spheres. 
A monolayer, formed by identical spheres in a hexagonal configuration, is placed on a tilted plane surrounded by a small fence that sustains the spheres, mimicking the disposal of fruits in the market. Then, a random continuous extraction process of spheres is imposed until the collapse. For this simple numerical experiment, a phase diagram was obtained to visualize the occurrence of avalanches triggered by vacancies as a function of the tilting angle, system size, and friction coefficient. More importantly, a sub-zone was found where we can predict the critical number of extractions until the avalanche takes place. The prediction is made from an evolution model of the average coordination number based on statistical considerations. The theoretical prediction also gives a constant critical void fraction of spheres, which implies the system collapses at a critical packing fraction.
 
\end{abstract}

\maketitle

%%%%%%%%%%%%%%% Introduction

\section{\label{sec:Intro}Introduction}

Granular avalanches are large-scale phenomena that can be triggered at small scales \cite{Daerr:1999hf,Lastakowski:2015jo,Salinas:2021ib}. However, the effectiveness of a small-scale precursor (for example, a particle hitting another after falling a small distance) depends on its ability to produce a chain effect that collectively breaks the strength of the entire system. This duality between small-scale precursors and their large-scale effect is still an open question. Some progress has been made in revealing the small-scale phenomenology by imposing shear through the system's walls \cite{Daniels2008,LeBouil:2014hp,Bares2017,Lherminier-Ramos2019}. This forcing corresponds to a large-scale perturbation imposed instantaneously to the whole system, as it is also when inclining, rotating, or shaking the entire granular assembly. Motivated by everyday phenomena, we want to see how random-small-scale perturbations progressively cause the loss of global stability, triggering large-scale avalanches.

\begin{figure}[!b]
\includegraphics[width=1\columnwidth]{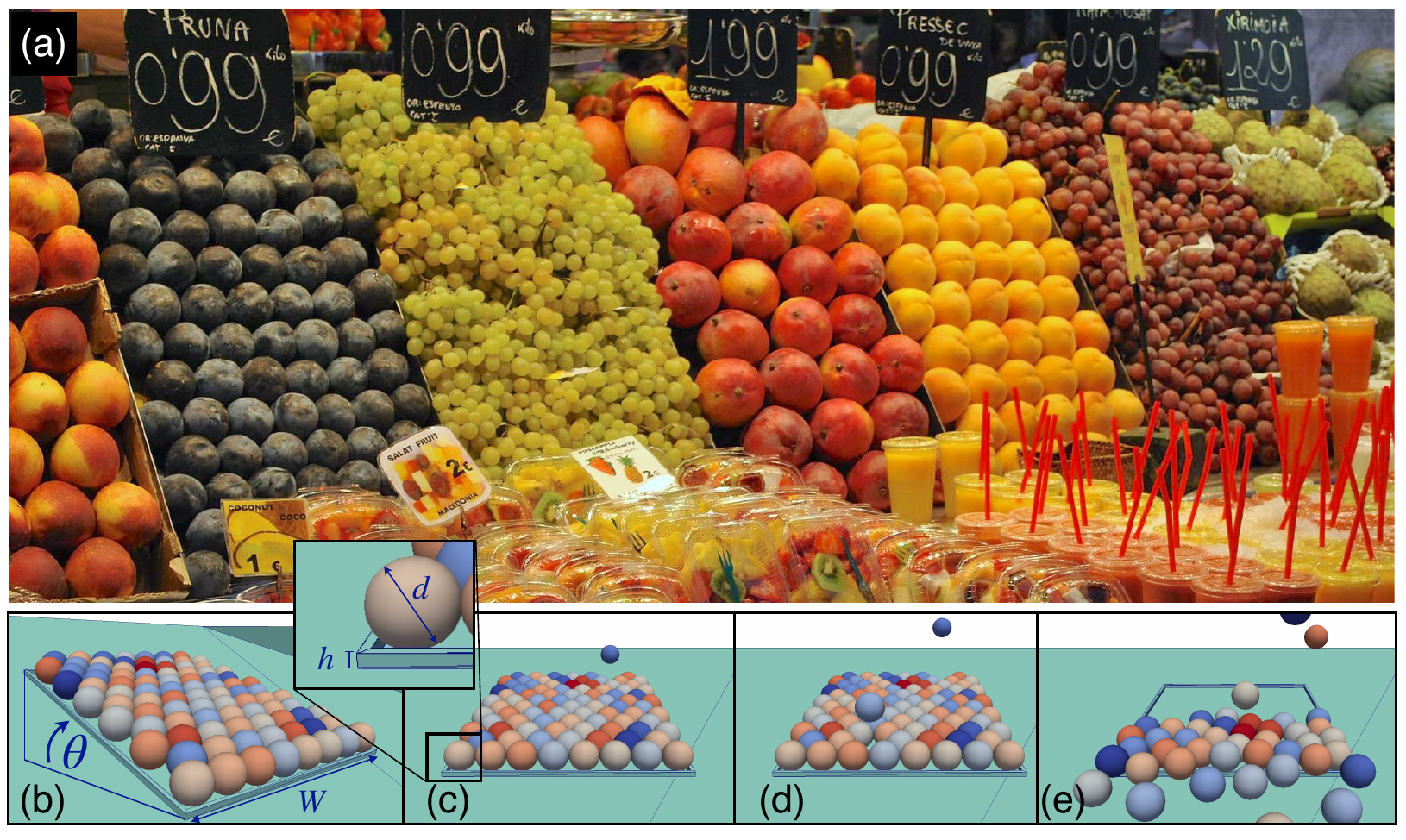}% Here is how to import EPS art
\caption{\label{fig:wide} Avalanches triggered by random extractions. (a) Arrangement of fruits in the market (photo credits in \cite{FotoCredit}). The bottom panels show a numerical simulation for a system with 90 spheres, randomly colored. (b) Presents a global view of the system, tilted with an angle $\theta$. In (c), the first sphere is extracted and the second in (d). In (e), an avalanche takes place after the extraction of 12 spheres. The inset in (c) highlights the fence bounding the system.}
\end{figure}

The example that inspired us is depicted in Fig. 1: vendors in the market stack the fruit trying to make most of it accessible to whoever wants to take it, which happens by placing the container with an optimal inclination. If the inclination is too slight, some fruit will be more difficult to reach. For a higher inclination, it will be easier to pick the fruit, but the stability of the stack could be compromised. Especially for this last case, one may ask how many fruits one can extract without producing the collapse of the pile (see Fig. 1). This simple system, familiar to everyone, presents two novelties compared to the literature on granular materials. First, fruits are generally displayed in hexagonal or square arrangements, which is possible because of their low size-dispersion. However, this perfectly crystallized configuration remains poorly studied \cite{German2022}. The second remarkable novelty corresponds to the way of triggering the avalanche. The generation of vacancies by fruit extractions corresponds to a new precursor of avalanches, producing both local rearrangements and a progressive decline in the coordination number, which weakens the contact structure globally.
Intuitively, such a system should be ruled by the contact structure between neighbors, determining forces distribution and system's rigidity, and, more broadly, by the system's packing fraction \cite{Lumay:2005ug, Scheller:2006gs, Dorbolo:2011kl,GravishEtAl2010,GravishGoldman2014}.

A disordered monolayer of particles placed on a tilted plane was considered before by Dorbolo \cite{Dorbolo:2005jp}: the slope of the plane is increased quasi-statically until a critical angle, where the stack becomes unstable and produces an avalanche. In some cases, only the lower part of the particles falls, while the rest remains in place. This observation represents a manifestation of the Jansen effect, revealing the complex network of contacts between spheres \cite{Dorbolo:2005jp}. Contact networks in granular assemblies indeed show filamentary structures or force chains ruling the stability of the media \cite{Liu:1995gi, Jaeger:1996ut, Majmudar:2005bn, Wensrich:2021eu} and physically connecting micro and macro scales. For instance, a local failure could be extended in a long-ranged response through a force chain without producing significant changes in particles' positions or orientations \cite{Wyart2005, Bares2017}. Therefore, exploring the so-called mesoscale variables in granular avalanches is essential. Accordingly, we focused our attention on the coordination number $z$.

Finally, it should be noticed that in compaction processes, particularly in monolayers, an increase in the packing fraction is usually observed \cite{Lumay:2005ug, Scheller:2006gs, Dorbolo:2011kl}. %\sout{However, let us suppose the compaction process triggers a flowing state. %like an avalanche. In this case, it generally does through a slight decrease of the packing fraction just before the avalanche, related to Reynolds dilatancy \cite{andreotti2013granular}}. 
However, in cases where a compaction process triggers a flowing state,  it generally goes through a slight decrease of the packing fraction just before the avalanche, related to Reynolds dilatancy~\cite{andreotti2013granular}. Indeed, Levy dit Vehel and collaborators \cite{Vehel-Ramos2021} subjected a monolayer to shear by applying torque to a cylindrical film configuration, stressing the relationship between the magnitude of the global dilation and the intensity of catastrophic events. Consequently, it seems reasonable to examine dilation to forecast sudden events like avalanches.

%This article is organized as follows. Our model is presented in section \ref{sec:Model}, together with its numerical implementation. Our results are presented in section \ref{sec:Results}, particularly a phase diagram for avalanches and the study of the evolution of the coordination number and elastic energy in the system. Then, the regime where avalanches are ruled by the decrease of elastic energy is discussed, followed by the conclusions in Section \ref{sec:Conclusions}.

%%%%%%%%%%%%%%% Model & Implementation

\section{\label{sec:Model} Setup and numerical implementation}

We performed this study using the \textit{Molecular Dynamics}~\cite{cundall1979discrete} discrete element method (DEM), through the \textit{ViscElMat} module of YADE \cite{yade}. This technique allowed us to compute spheres' movements and interactions, giving detailed information about their contacts.

 The setup consists of a rectangular monolayer arrangement of identical spherical particles of diameter $d=\SI{4}{\cm}$ and mass $m=\SI{2.93}{\g}$, similar to table-tennis balls. These spheres, subjected to gravity in the vertical direction ($g=\SI{9.8}{\m/\s^2}$), are disposed on a tilted plane, with an angle $\theta$ with respect to the horizontal plane (see Fig. \ref{fig:wide}b). Initially, the spheres are crystallized forming a perfect hexagonal configuration. 
 The arrangement is sustained by a small fence placed across the monolayer boundary. The fence must be high enough to retain the spheres while allowing an avalanche to take place (for instance, fences higher than $0.5d$ set a very different problem). Therefore, we chose a height of $h=0.17d$ (see inset in Fig. \ref{fig:wide}c) to have a good compromise. We checked that slight variations from this specific value have little influence on our results, as discussed in Section \ref{sec:ParametricAssessment}.
 
Our smallest system corresponds to a rectangular box allowing a monolayer of $90$ spheres. Due to the hexagonal packing, horizontal rows alternate between eight and seven spheres in this case (see Fig. \ref{fig:wide}b). Its size can be described by its width $W$, which in this case corresponds to $8d = \SI{32}{\cm}$. We scaled both sides of the rectangular box by a factor from 1 to 5 to analyze size effects. Therefore, the width $W$ takes values of $8 d, 16 d, 24 d, 32 d$, and $40 d$, the box's height scales accordingly to preserve the box's aspect ratio, and the initial number of spheres $N_0$ becomes 90, 372, 846, 1512, and 2370, respectively.

To initialize the simulations, after the spheres are placed on the tilted surface, a gentle horizontal movement is applied to the whole system to settle the spheres in a stable position. 
The velocity $v(t)=v_0\sin(2\pi t/T)$ is applied to the system for $\SI{2}{\s}$, where $v_0=\SI{0.1}{\m/\s}$ and $T=\SI{1}{\s}$. During this initial process, the relative movement between spheres is negligible and the hexagonal configuration is preserved.

After initialization, a random extraction process starts, taking the spheres one at a time from inside the rectangle that contains and supports the spheres \footnote{For each extraction process, all the spheres in the system are labeled, and one of them is chosen according to a uniform random distribution.}. A velocity perpendicular to the inclined plane is imposed to each particle selected to be removed (see Figs. \ref{fig:wide}.c and \ref{fig:wide}.d). 
The interval between extractions was three seconds. As the characteristic time $\tau$ for a sphere to displace one diameter is near 0.1 s ($\tau \approx \sqrt{d/g\sin \theta}$), there is enough time to allow spheres to relocate to their new stable position or escape outside the fence. By doing so, the quasi-static evolution of the system is better described by the number of spheres extracted rather than time.

The Molecular Dynamics method uses particle-particle and particle-plane contact forces when the elements overlap on $\delta$ \cite{luding2008introduction}. The model for the normal contact force $F_n$ is viscoelastic in its linear form: $F_n=k_n\delta+c_n\dot{\delta}$, where $k_n$ and $c_n$ are constants. The tangential contact force $F_t$ has the same form with constants $k_t$ and $c_t$, added to Coulomb friction with the same static and dynamic friction coefficient $\mu$. The constant $k_n$ for the normal force is set to $\SI{10,000}{\N/\m}$ to obtain an overlap $\delta<10^{-3}d$ in all the simulations performed. In Section \ref{sec:ParametricAssessment}, we verified that changes on $k_n$ have little influence on the results except when the spheres are highly rigid ($k_n>\SI{100,000}{\N/\m}$), as contacts are easily lost because of the low overlap between particles.
The first constant in the tangential force is fixed to $k_t = 0.5k_n$, as used in numerous DEM works (see, for instance, \cite{Rojas:2019ia}). The dissipative terms are fixed to obtain a restitution coefficient $e=0.7$. For most simulations, we set the Coulomb friction coefficient to $\mu  = \tan(35^{\circ}) = 0.7$. However, we explore the dependence on $\mu$ in sections \ref{sec:ModelGlobalResults} and \ref{sec:ParametricAssessment}.

\begin{figure}%[b]
\includegraphics[width=1\columnwidth]{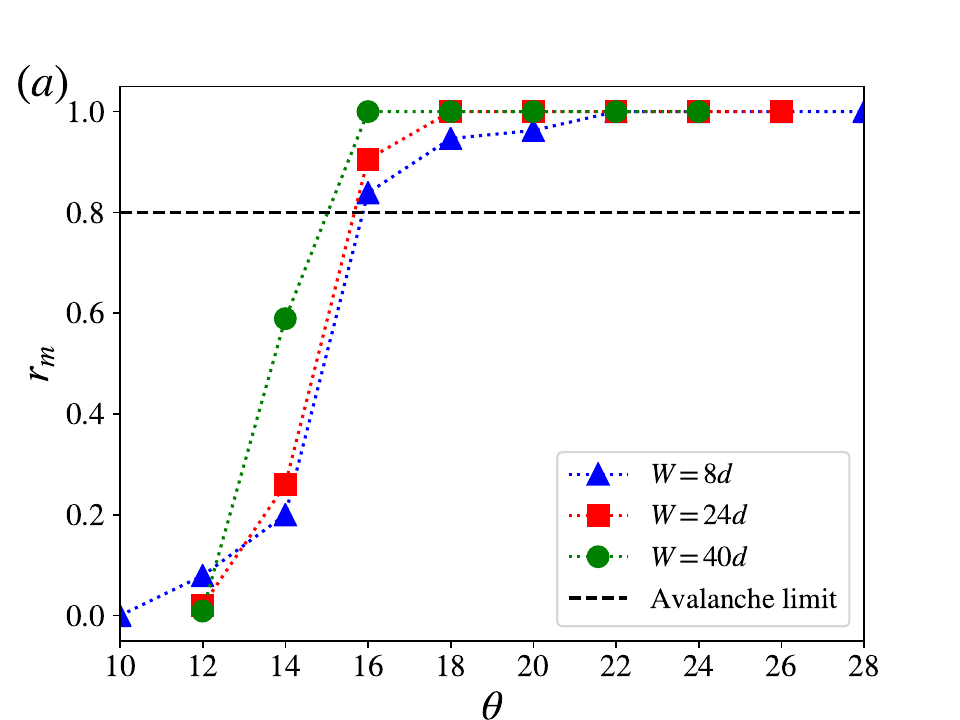} 
\includegraphics[width=1\columnwidth]{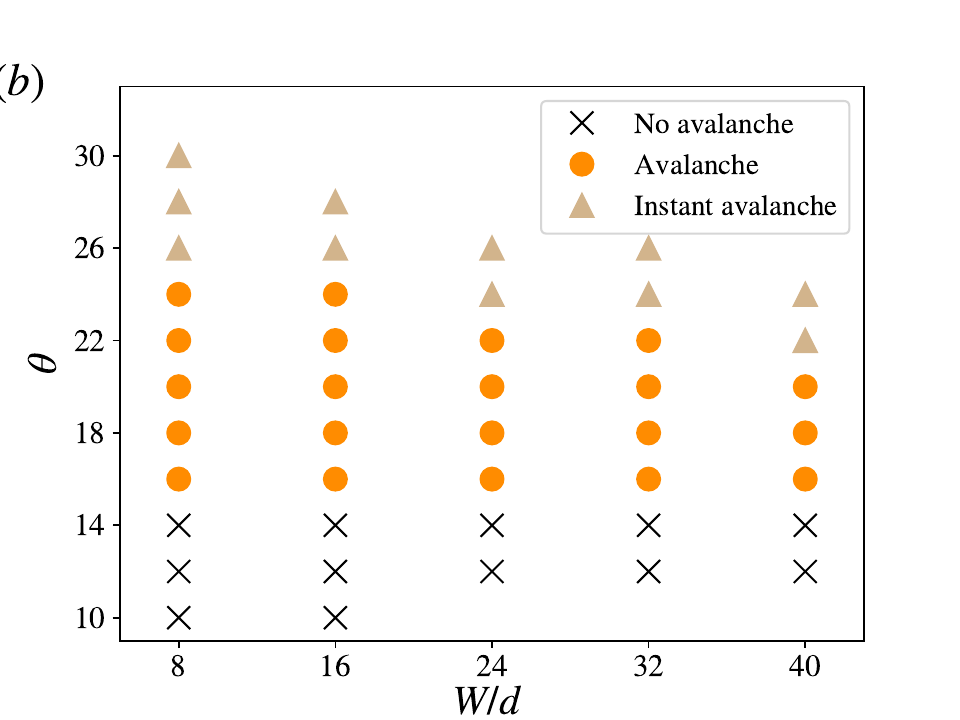}
\includegraphics[width=1\columnwidth]{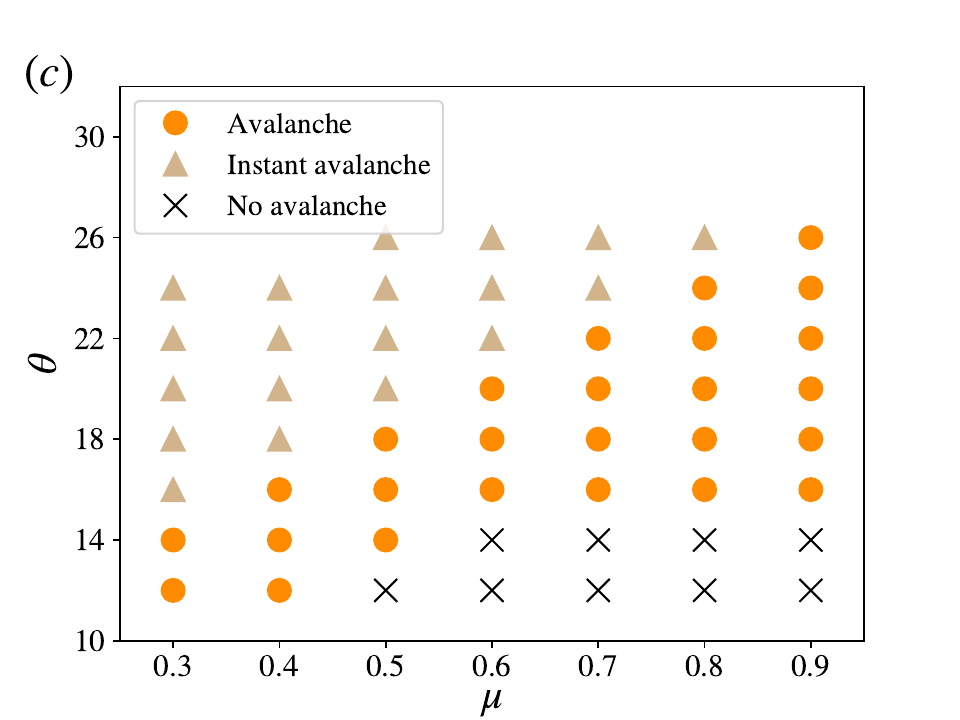}
\caption{\label{fig:GlobalResults} Phase diagrams for avalanches. (a) Migration ratio $r_m$ of spheres as a function of the inclination angle. Symbols represent different system sizes. The horizontal dashed line at 0.8 shows our threshold to define an avalanche. (b) and (c) show phase diagrams with the three main regimes observed: No avalanche (crosses), Instant avalanche (triangles), and Avalanche (circles). (b) Phase diagram of angle $\theta$ and dimensionless system's width $W/d$, for a fixed $\mu$ = 0.7. (c) Phase diagram of angle $\theta$ and friction coefficient $\mu$, for a fixed $W/d$ = 24.}
\end{figure}

%%%%%%%% Results

\section{\label{sec:Results} Results}

\subsection{\label{sec:ModelGlobalResults} Phase diagram for avalanches}

To decide whether an avalanche occurs or not, we consider two quantities: the number of spheres inside the system $N_{in}$; and the spheres that migrate outside the system immediately after a given extraction, $N_{out}$. Both quantities depend on the \textit{extraction number} $n$: the umpteenth sphere extracted from the system. 
The parameter defining the \textit{migration ratio} $r_m$ is:  \begin{equation}
    r_m= \max{\left(\frac{N_{out}(n)}{N_{in}(n)}\right)},
\end{equation} were the maximum is taken along the whole extraction sequence.
We consider that $r_m> 0.8$ constitutes an avalanche. Figure \ref{fig:GlobalResults}a presents the ratio $r_m$ as a function of $\theta$ for three system sizes ($W=8d,24d,40d$), where we draw a dashed line for the avalanche criterion. 
Figure \ref{fig:GlobalResults}a shows that for a fixed angle $\theta$, in general, bigger systems imply bigger spheres migrations (bigger $r_m$ values). 

Defining $n_{c}$ as the extraction number where the avalanche criterion is attained, we built phase diagrams for the tilting angle $\theta$ as a function of the dimensionless system's width ($W/d$, Fig. \ref{fig:GlobalResults}b) and friction ($\mu$, Fig. \ref{fig:GlobalResults}c). We distinguished three phases:
\begin{itemize}
\item \textit{no avalanche} zone, where $r_m\leq 0.8$ ($\times$ markers), and all the spheres can be extracted without the collapse ($n_{c}~\to~N_0$);%\infty$);
\item \textit{instant avalanche} zone, where $r_m=1$ and $n_{c}~=~0$ (triangle markers); and,
\item \textit{avalanches triggered by extraction of spheres}, where $0.8<r_m\leq 1$ and $n_{c}>0$ (orange circle markers). This case is the main focus of our study. 
\end{itemize}
It can be noticed in the upper part of Fig. \ref{fig:GlobalResults}b that instant avalanches start at smaller angles for larger systems (larger $W/d$): this means that boundaries play a significant role for smaller systems. On the other side, for low angles, there is a wide range where we can extract every sphere from the system without producing an avalanche. Therefore, there is a critical angle setting an inferior limit for avalanches, $\ang{16}$ in this case. Intuitively, this happens because of a stronger influence of the friction with the floor. 
Indeed, Fig. \ref{fig:GlobalResults}c shows that reducing the friction coefficient $\mu$ makes the limiting angle smaller. Figure \ref{fig:GlobalResults}c, at large angles, shows that $\mu$ significantly modifies the threshold for instant avalanches: increasing friction between neighboring spheres globally stabilizes the system, enlarging the region where the extraction process triggers the avalanche (shown as orange circles).

\begin{figure*}[t!]
\includegraphics[width=2\columnwidth]{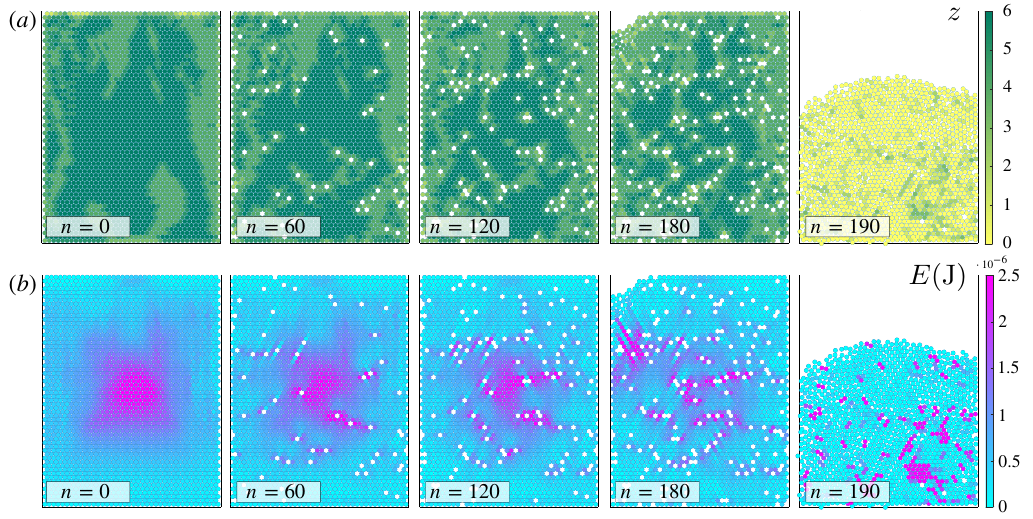}% Here is how to import EPS art
\caption{\label{fig:wide2} Spatial distribution of coordination number and elastic energy. (a) Coordination number field $z$, and (b) elastic energy field $E$. Initially, the system contains 2370 spheres ($W$ = 40$d$) and the tilting angle is fixed to $\theta = \ang{20}$. Five states are presented by means of snapshots obtained after the extraction of $n$ particles. From left to right, are shown $n$~=~0, 60, 120, 180 and 190. In the last step, the collapse is taking place, as shown in more detail in Fig. \ref{fig:wide3}. A movie of the whole extraction process can be found in \cite{SuppMat}.}
\end{figure*}

\begin{figure*}[t!]
\includegraphics[width=2\columnwidth]{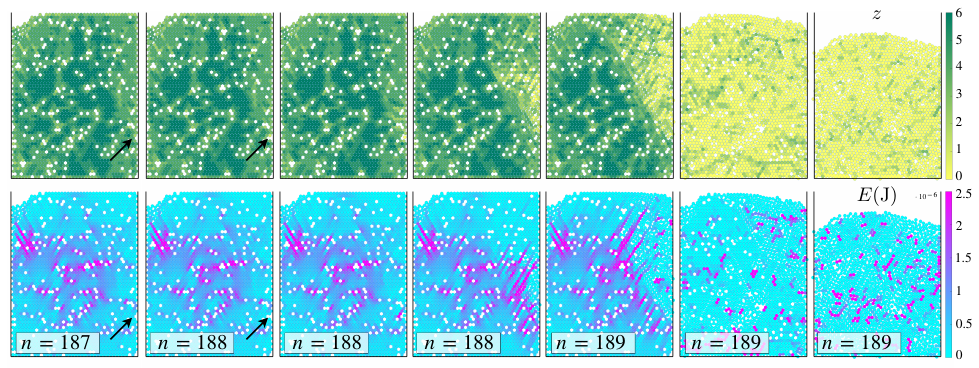}% Here is how to import EPS art
\caption{\label{fig:wide3} Temporal evolution during the collapse. As in Fig. \ref{fig:wide2}, we show the coordination number and elastic energy after the sphere's extraction $n$~=~188, indicated with a black arrow. Color coding is the same as in panels (a) and (b) from Fig.  \ref{fig:wide2}. A movie of the whole extraction process can be found in \cite{SuppMat}.}
\end{figure*}

%%%%%%%%%%%%%%% Specific Results

\subsection{\label{sec:SpecificResults} Evolution of the system}

In order to get a qualitative view of the processes involved, it is helpful to focus on the coordination number $z$ and the elastic energy of the contacts $E$, obtained from simulations. The coordination number was computed by counting lateral contacts sphere-sphere and sphere-fence (i.e.: contacts with the bottom plane are excluded), which gives 6 for spheres in the bulk of a 2D, densely packed hexagonal configuration.
The elastic energy was computed in each contact as $1/2k_n\delta^2$. 
We present examples of fields $z$ and $E$ in Fig. \ref{fig:wide2}, Fig. \ref{fig:wide3}, and in the supplemental movie \cite{SuppMat}. 

\begin{figure*}[t]
\includegraphics[width=1\columnwidth]{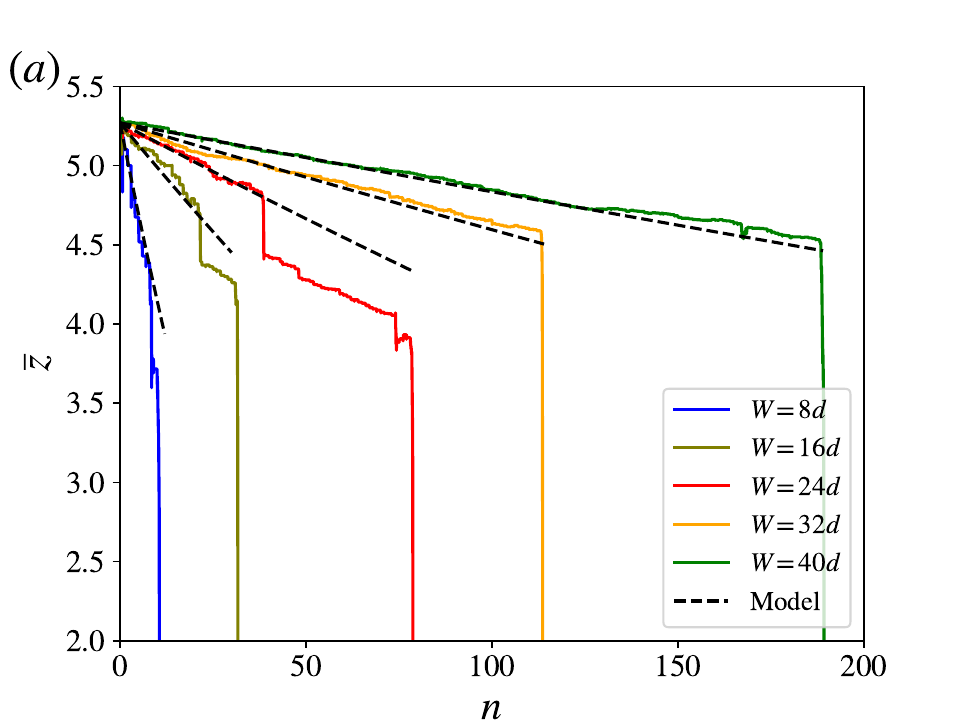}
\includegraphics[width=1\columnwidth]{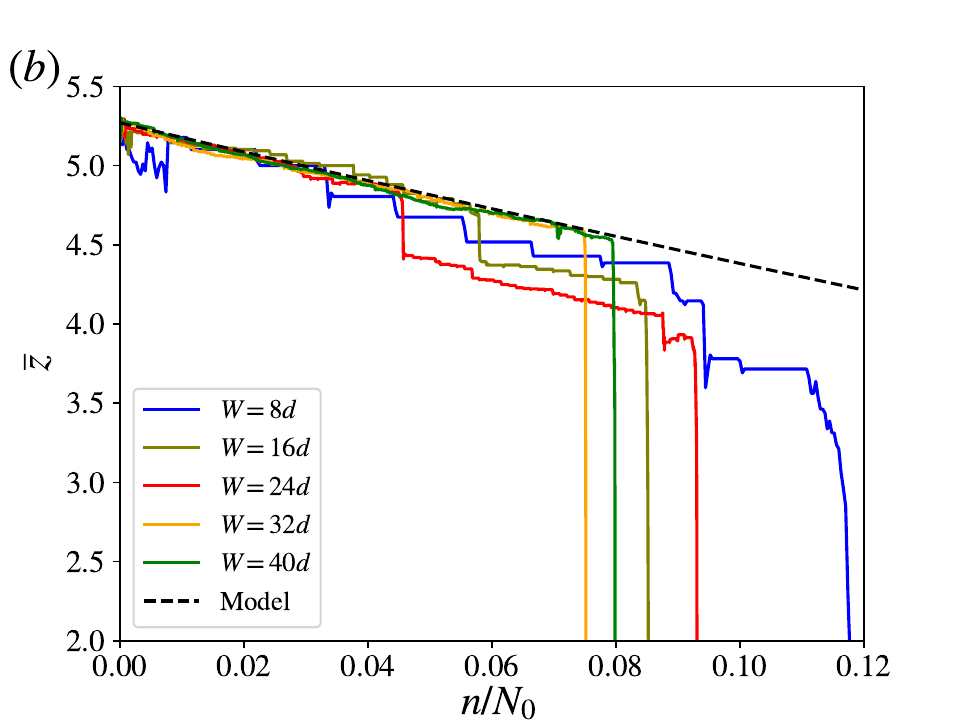}
\caption{\label{fig:Z_E} Evolution of averaged coordination number. (a) Coordination number $\bar{z}$ is presented as a function of the extraction number $n$, for a fixed tilting angle $\theta=\ang{20}$ and different system's sizes, specified by the box's width $W$. (b) Coordination number as a function of the extraction number normalized by the initial particle number $N_0$. In both panels, segmented lines represent the model given in equation \eqref{eq_zn2}.}
\end{figure*}

Figure \ref{fig:wide2}a shows the coordination number $z$ of the system for $W/d=40$ and $\theta=\ang{20}$. 
This figure corresponds to a sequence for different extraction numbers ($n$ values) from the system's initial state until it collapses. For the initial condition ($n=0$), there is a considerable amount of particles with less than 6 contacts. 
However, this loss of contacts does not indicate a departure from the hexagonal configuration. Instead, it comes from fewer contacts at the systems' boundaries and some losses during system initialization. Also, for $n=0$, Fig. \ref{fig:wide2}b shows that the elastic energy $E$ is not homogeneous but more concentrated in a small central core, which implies a non-hydrostatic pressure distribution. This non-hydrostatic distribution is consistent with the Janssen effect, where the strongest pressure is not located at the bottom of the medium but in a central place. This effect has been reported for confined granular monolayers \cite{Dorbolo:2005jp,Scheller:2006gs}. Then, for $n=60$ and $120$ the coordination number decreases in magnitude across the system by forming channels of lower contacts. The elastic energy also decreases, particularly in the central core. At $n=180$ an empty zone appears in the upper-left corner of the system produced by a small migration of particles. Note that this migration does not constitute an avalanche because the avalanche criterion $r_m>0.8$ is not satisfied. A consequence of this departure is an increment of the elastic energy just under the empty zone, producing diagonal segments of strong elastic energy, which increases the system's strength locally. For $n_c$ = 190 the system collapses and all particles depart outside the fence of the system. 
At this point, most of the contacts are lost (low $z$ values), and consequently, the elastic energy is globally reduced, except for some intense interactions between particles.

Figure \ref{fig:wide3} presents a close-up of the collapse, offering insights into the loss of stability of the whole system. We focus on the extraction at $n=188$ (indicated by black arrows on the left panels), where the collapse started. Here the snapshots are consecutive and also appear at the end of the supplemental movie \cite{SuppMat}.
In this case, the critical sphere (indicated by arrows) was extracted incidentally from a place close to the wall. Then, some spheres fell by the side, leaving space for a slide to develop, destabilizing the whole system. The slide can be noticed particularly in the plots of $z$ as a yellow region (low $z$) that grows with time. In other words, the loss of a sphere that acted as a support for others (structural support) allowed more spheres to move, creating a region with a strong shear that produced the system's collapse.

Figure \ref{fig:Z_E} presents a picture complementary to the previous discussion. It shows the average coordination number of the system $\bar{z}$ as a function of the extraction number $n$, for different system sizes $W$, and considering a fixed angle $\theta=\ang{20}$. Regardless of $W$, the coordination number starts near $5.25$, less than 6 for a hexagonal configuration, as discussed earlier. 
$\bar{z}$ decay monotonously with $n$ until a final sharp drop when the avalanche is produced. All curves follow a trend of linear decay before the avalanche, except for two punctual intermediate drops for $W=16d$ and $24d$ associated to internal rearrangements where the hexagonal configuration is lost locally. The dashed lines correspond to equation~\eqref{eq_zn2}, coming from a simple statistical model to be described next.  
Figure \ref{fig:Z_E}b shows $\bar{z}$ as a function of the void fraction $n/N_0$, where $N_0$ corresponds to the initial number of particles in the system. All curves collapse near the model, reinforcing the idea of a statistical rule dominating the evolution of $\bar{z}$. In the end, when the avalanche occurs, the granular medium flows and becomes less dense, which causes a fall in all $\bar{z}$ curves. From Fig. \ref{fig:Z_E}b, we can notice that in all cases, the void fraction is similar at collapse and near a 10\%. 

\subsection{\label{sec:SpecificResultsRules} Statistical rule for the evolution of the average coordination number $\bar{z}$}

To formulate a simple statistical model for the average coordination number, we will consider that the lost contacts after an extraction $n$ are homogeneously distributed between the $N^n_{in}$ spheres remaining in the system. Note that this hypothesis implies an equal and, in general, non-integer number of contacts for all particles. Denoting the total number of contacts in the system by $C_n$, then considering that each contact acts in two spheres, the average coordination number is written:
\begin{eqnarray}
    \bar{z}_n &=& \frac{2C_n}{N^n_{in}}.
\end{eqnarray}
Assuming that the contacts in the system remain the same aside from the zones of extraction, the contacts $C_n$ are equal to the contacts for the extraction $n-1$, associated to the $N^n_{in}$ particles remaining in the system (which are $\bar{z}_{n-1}N_{in}^n/2$), less the lost contacts around the new vacancy ($\bar{z}_{n-1}$):
\begin{eqnarray}
    \nonumber
    \bar{z}_n %&=& %\frac{2\left({C_{n-1}-\bar{z}_{n-1}}\right)}{N^n_{in}}\\
    %\nonumber
    &=& \frac{2\left({\bar{z}_{n-1}N_{in}^n/2-\bar{z}_{n-1}}\right)}{N^n_{in}}\\
    &=&\bar{z}_{n-1}\left({1-\frac{2}{N^n_{in}}}\right).
    \label{eq_zn1}
\end{eqnarray}

The equation~(\ref{eq_zn1}) can be written as a function of the initial average coordination number $\bar{z}_0$ and the initial number of spheres inside the system $N^0_{in}=N_0$:
\begin{equation}\label{eq_zn2}
    \bar{z}_n = 
    \begin{cases}
        \bar{z}_0&\text{, } n=0\\
        \bar{z}_0 \displaystyle \prod_{i=1}^{n}\left({1-\frac{2}{N_0-i}}\right)& \text{, } n=1,2,3,\ldots,
    \end{cases}
\end{equation}
where the factor $2/(N_0-i)$ % $\frac{2}{N_0-i}$ 
represents the decreasing fraction of the average coordination number when the $i$-th particle is extracted from the system. 
We included the results of this model as dashed lines in Fig. \ref{fig:Z_E}. Strikingly, the statistical rule is satisfied most of the time, implying that the general hexagonal configuration is mostly preserved, except in the vacancies themselves. 

\begin{figure*}%[h!]
    \centering
    \includegraphics[width=1\columnwidth]{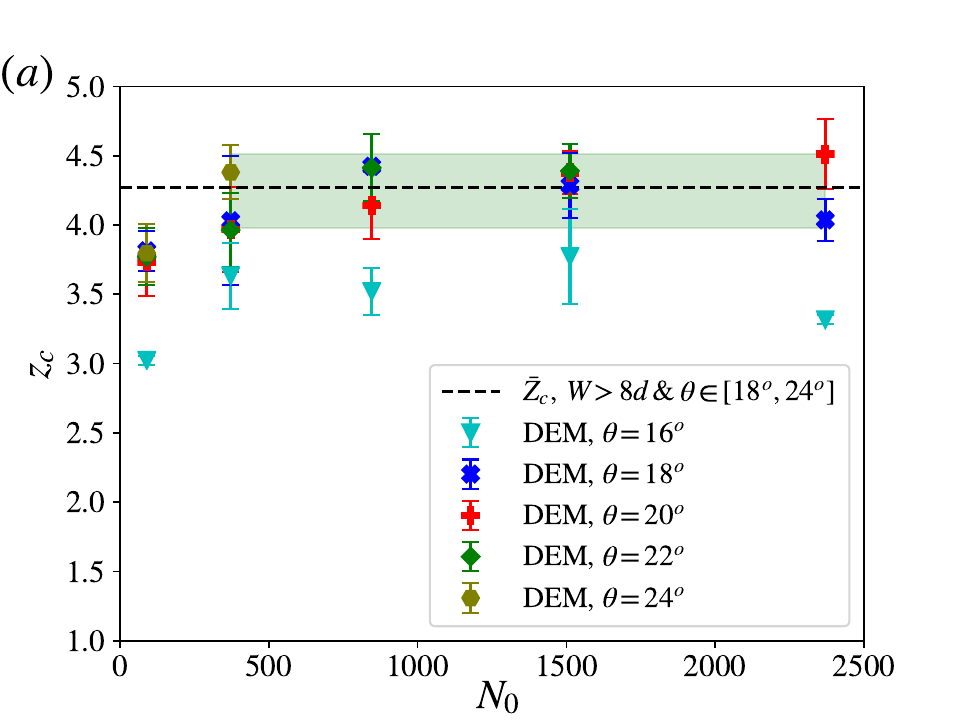}
    \includegraphics[width=1\columnwidth]{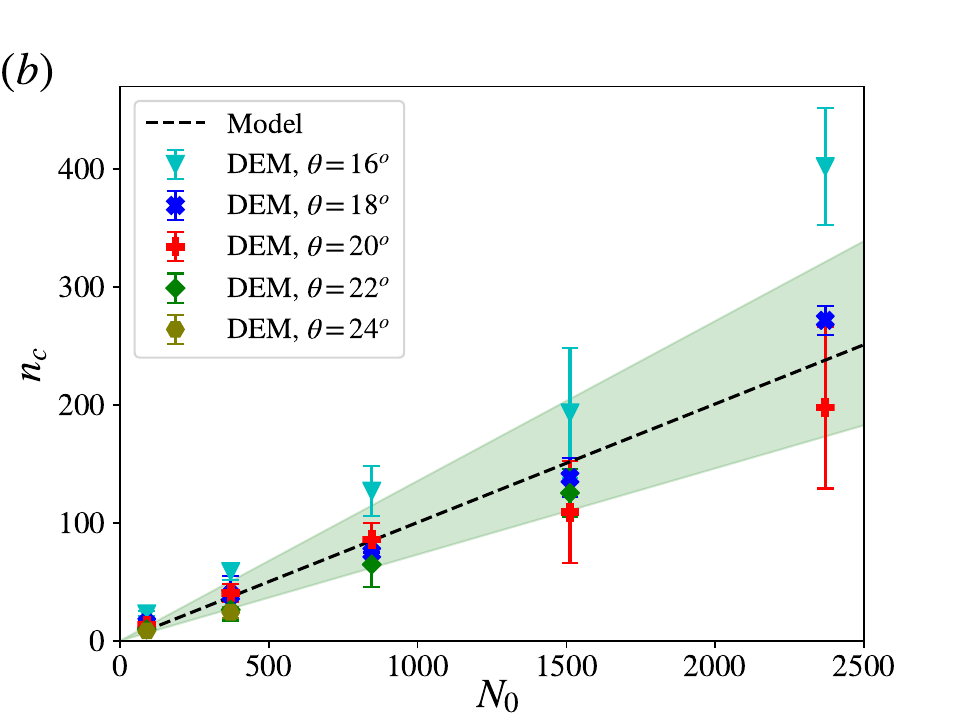}
    \caption{Critical coordination number $\bar{z}_c$ (a) and extraction number $n_c$ (b) as a function of the initial number of spheres $N_0$. Every point is obtained by averaging three simulations (each with a different random extraction sequence), and the error bars correspond to standard deviations. Colors and symbols are associated with the tilting angles used. The shaded area in (a) and (b) was added for showing the  dispersion of the data. In (a), the dashed line indicates $\bar{z}_c$ = 4.27, obtained as the mean value between all points with $N_0 > 90$ and $\theta>18^{\circ}$.  The shaded area  represents the maximum fluctuation of the coordination number for $N_0 > 90$ and $\theta>18^{\circ}$ ($\bar{z}_{min}=3.98$ and $\bar{z}_{max}=4.51$). In (b), the dashed line corresponds to the model in equation~\eqref{eq_nav}, using $\bar{z}_c$ = 4.27 and $\bar{z}_0$ = 5.22. $\bar{z}_0$ was obtained from averaging on the set of simulations. The shaded area represents the models' fluctuation using $\bar{z}_c$ between $\bar{z}_{min}$ and $\bar{z}_{max}$ from (a).} %Fig. \ref{fig_avalanche_prediction}a.}}
    \label{fig_avalanche_prediction}
\end{figure*}

%%%%%%%%%%%%%%% Critical extraction number

\subsection{\label{sec:Critic_extraction_nunmber} Critical extraction number $n_c$}
%\paragraph{Prediction for critical energy and packing fraction. \label{sec:Critic_extraction_nunmber}}

The model given by equation~\eqref{eq_zn2}
implies a continuous loosening process, where the extraction of the $n$-th particle produces a drop in the average coordination number $2/(N_0-n)$. %, of $2/(N_0-n)\bar{z}_{n-1}$. 
This process ends at the extraction number $n_c$, when the coordination number reaches a critical value $\bar{z}_c$, where the avalanche is triggered. From equation~\eqref{eq_zn2} considering $i\ll N_0$, we obtain:
\begin{equation} \label{eq_nav}
    n_c = \frac{\ln\left({\displaystyle\frac{\bar{z}_c}{\bar{z}_0}}\right)}{\ln\displaystyle\left({1-\frac{2}{N_0}}\right)},
\end{equation}where $\bar{z}_0$ is the initial value of the average coordination number. Both $\bar{z}_0$ and $\bar{z}_c$ can be obtained from simulations.
Indeed, we obtain $\bar{z}_0$ = 5.22, and in Fig. \ref{fig_avalanche_prediction}a, we present results of $\bar{z}_c$ for a wide range of parameters. Most data attain a plateau with a nearly constant $\bar{z}_c$. The only exceptions are cases with $W = 8d$ with strong border effects and the angle $\theta  = \ang{16}$, which is very close to a transition according to Fig. \ref{fig:GlobalResults}b. This angle sets the validity limits for the model. We obtain $\bar{z}_c = 4.27$ excluding those cases.

With the ratio $\bar{z}_c/\bar{z}_0$, we model how $n_c$ grows with $N_0$, as presented in Fig. \ref{fig_avalanche_prediction}b.
We obtained fair agreement between the prediction of $n_c$ and the numerical simulations. Also, simulations confirm that for most angles, the critical extraction number is $\theta$-independent, as predicted by our model. The independence on $\theta$ is striking yet natural: if the tilting angle increases, the system is more susceptible to collapse, but at the same time the contact structure becomes stronger.

%%%%%%%%%%%%%%% Packing fraction

\subsection{\label{sec:Critical packing fraction} Critical packing fraction}

The model proposed in  Eq. (6) indicates that $n_c$ is a linear function of $N_0$ when it is large, in agreement with our DEM simulations (see Fig.~\ref{fig_avalanche_prediction}b). By fitting the model to a straight line, we obtain $n_c/N_0=0.10$. The quotient $n_c/N_0$ represents a critical void fraction which in turn gives us a constant critical packing fraction $\phi_c$:

\begin{eqnarray}
    \phi_c &=& \phi_{HE}\left({1-\frac{n_c}{N_0}}\right)=0.82, \label{eq_phic}
\end{eqnarray}
where $\phi_{HE}=\pi/(2\sqrt{3}) \approx 0.907$ represents the packing fraction of the hexagonal configuration for a 2D monolayer of disks. 
A possible interpretation of $\phi_c$ is that subtracting particles is analogous to dilating the system continuously. In this context, $\phi_c$ corresponds to a threshold for the medium to flow. Indeed, a similar value of $\phi_c=0.81$ is found for DEM simulations of disks subjected to simple shear~\cite{dacruz2005rheophysics}. In this sense, the system starts crystallized at a packing fraction of $\phi_{HE}\approx 0.907$, then passes through the \textit{Random Close Packing Fraction} $\phi_{RCP}\approx 0.853$ \cite{blumenfeld2021disorder}, finishing at the collapse where the system flows at $\phi_c=0.82$. This might suggest that the avalanches in our system are triggered by an induced shear due to particle extractions.

\begin{figure*}
    \centering
    \includegraphics[width=1\columnwidth]{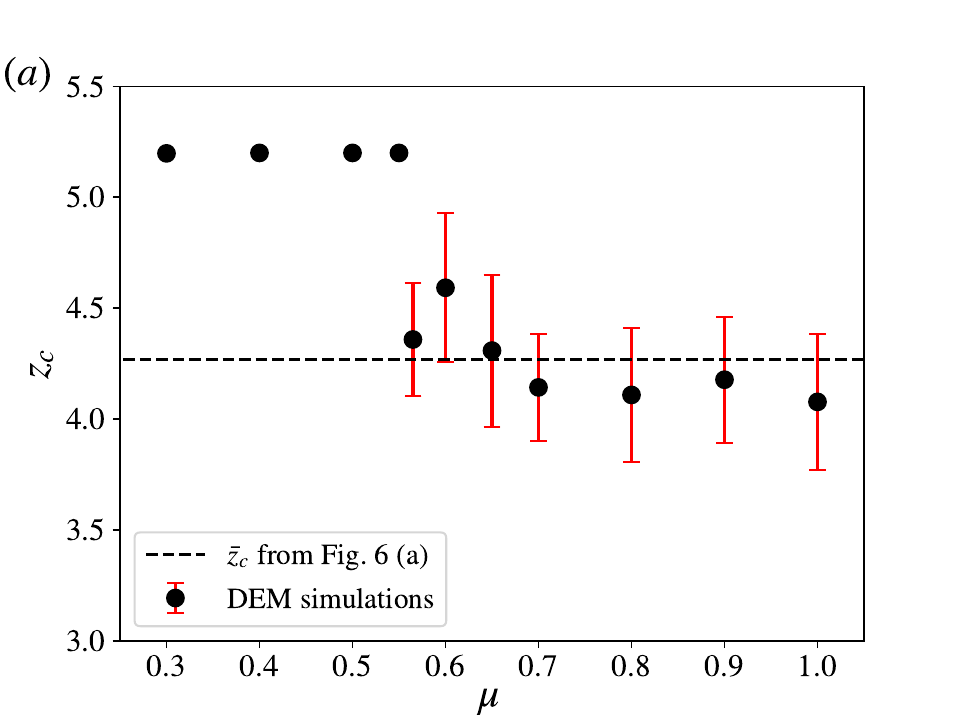}
    \includegraphics[width=1\columnwidth]{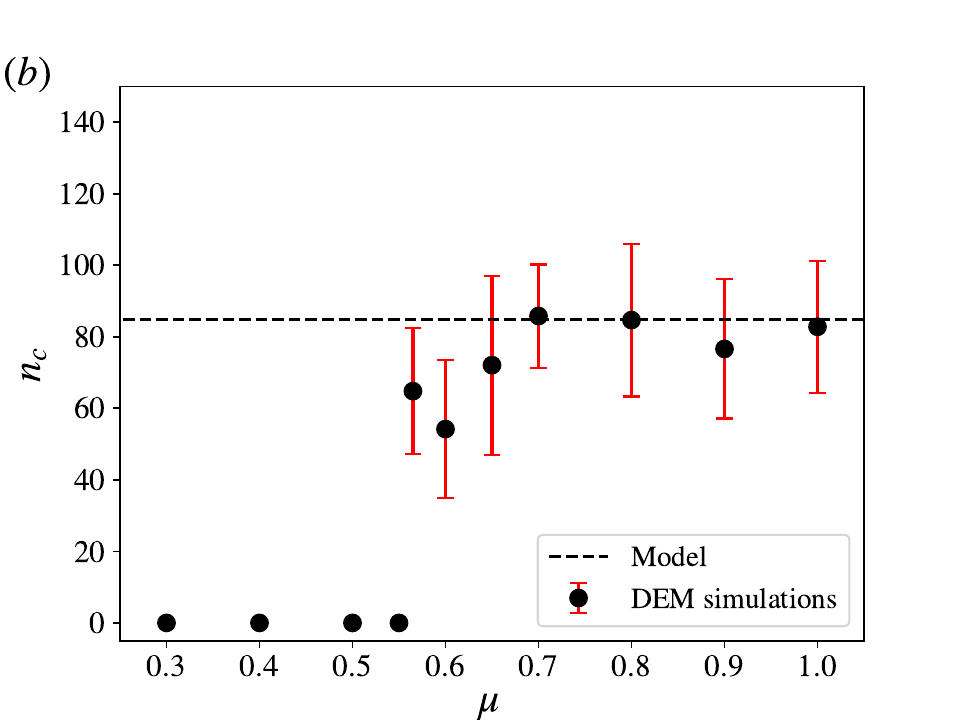}
    \caption{Effect of modifying the friction coefficient between spheres (for $W/d=24$ and $\theta=\SI{20}{\degree}$). Each point represent the mean value of 15 simulations and the error-bars are the standard deviation. (a) Critical coordination number as a function of the friction coefficient. The horizontal dashed line represents the mean critical coordination number $\bar{z}_c=4.27$ obtained from Fig. \ref{fig_avalanche_prediction} (a). (b) Critical extraction number as a function of the friction coefficient. The horizontal dashed line represents the critical extraction number $n_c=84.9$ obtained from the proposed model (Eqn. \eqref{eq_nav} for $N_0=846$).}
    \label{fig_mu}
\end{figure*} \begin{figure*} %[h]
    \centering
    \includegraphics[width=1\columnwidth]{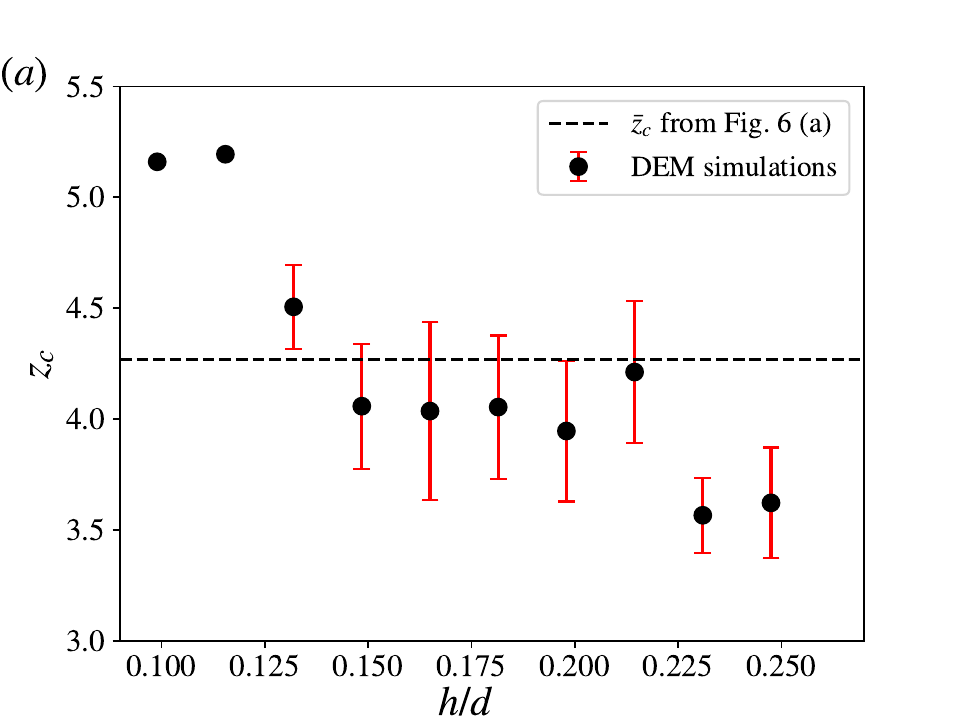}
    \includegraphics[width=1\columnwidth]{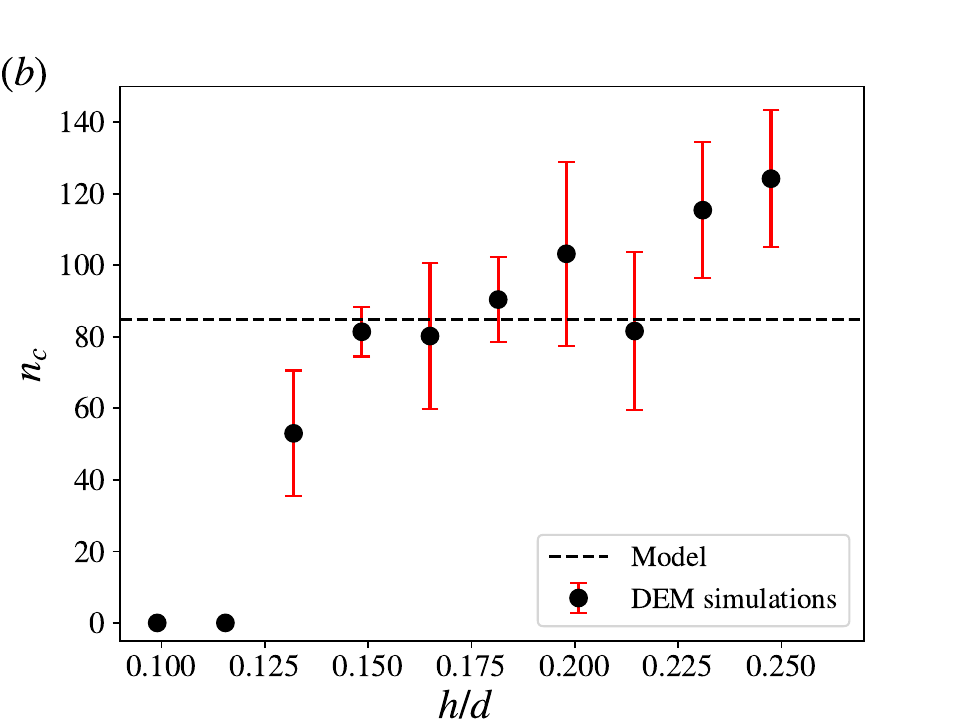}
    \caption{Effect of the fence height (for $W/d=24$ and $\theta=\SI{20}{\degree}$). Each point represent the mean value of 10 simulations and the error-bars are the standard deviations. (a) Critical coordination number as a function of the fence height. (b) Critical extraction number as a function of the fence height. Horizontal dashed lines are the same as in Fig. \ref{fig_mu}.}
    \label{Fence_height}
\end{figure*} \begin{figure*}%[h]
    \centering
    \includegraphics[width=1\columnwidth]{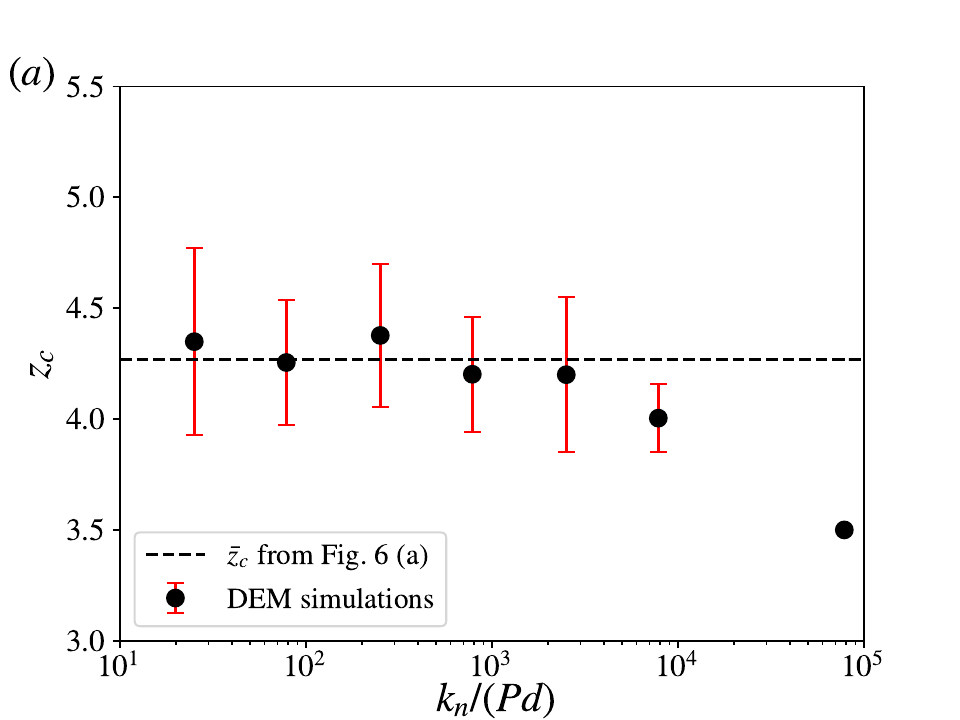}
    \includegraphics[width=1\columnwidth]{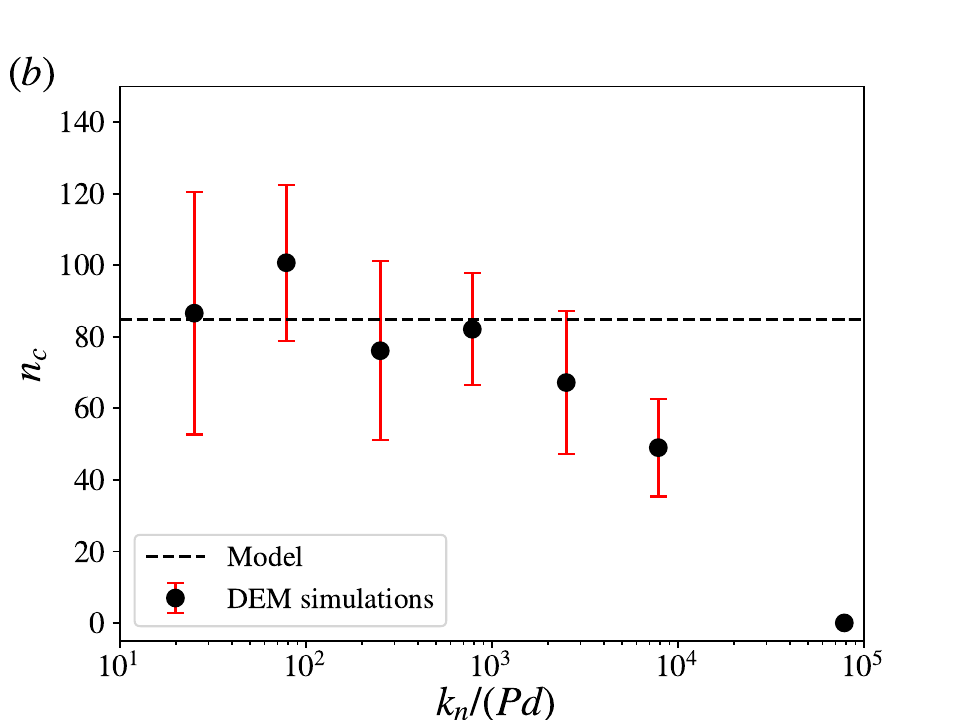}
    \caption{Effect of the spheres' stiffness (for $W/d=24$ and $\theta=\SI{20}{\degree}$). Each point represent the mean value of 10 simulations and the error-bars are the standard deviations. (a) Critical coordination number as a function of the dimensionless stiffness of the spheres. (b) Critical extraction number as a function of the dimensionless stiffness of the spheres. Horizontal dashed lines are the same as in Fig. \ref{fig_mu}.}
    \label{Fig_Stiffness}
\end{figure*}

%%%%%%%%%%%%%%% Influence of $\mu$.

\subsection{\label{sec:ParametricAssessment} Parametric study of $\bar{z}_c$ and $n_c$}

The results from Fig. \ref{fig:wide2} to Fig. \ref{fig_avalanche_prediction} were all obtained for a fixed friction coefficient $\mu = 0.7$. However, we already showed that friction influences avalanches, as shown in the phase diagram in Fig. \ref{fig:GlobalResults}c. Accordingly, we ask ourselves if friction influences the scenario for avalanches triggered by vacancies, as described by $\bar{z}_c$ and $n_c$. To answer the question, we performed the simulations shown in Fig. \ref{fig_mu}, with fixed $W = 24d$, $\theta=\ang{20}$ and varying $\mu$ in a wide range.

In the phase diagram of Fig. \ref{fig:GlobalResults}c, we always have instantaneous avalanches for $\theta=\ang{20}$ and $\mu < 0.6$. Therefore, $\bar{z}_c$ stays at its initial value in those cases. Indeed, Fig. \ref{fig_mu}a shows that $\bar{z}_c$ = 5.24 for $\mu < 0.55$. There is a jump for $\mu>0.55$, where $\bar{z}_c$ takes values around 4.27, the same obtained in Fig. \ref{fig_avalanche_prediction}a and indicated with a dashed line. Similarly, Fig. \ref{fig_mu}b shows that for $\mu$ larger than 0.55, $n_c$ exhibits only slight variations around the value obtained from the model \eqref{eq_nav}. In summary, while friction sets whether an avalanche occurs or not (Fig. \ref{fig:GlobalResults}c), it has a weak effect on $\bar{z}_c$ and $n_c$ in the range where vacancies trigger avalanches.

Similarly, we could test the influence of the height $h$ of the fence we used to retain spheres, as shown in Fig. \ref{Fence_height}. It is reasonable to expect some dependence on $h$ as the barrier becomes harder to overcome when $h$ increases. Indeed, Fig. \ref{Fence_height}(b) shows that $n_c$ increases linearly with $h/d$ in almost the whole range explored. Around $h/d$ = 0.17 (the value used in previous sections), variations in $n_c$ are only slight, as is the case for $\bar{z}_c$ in a broader range. 
For values of $h/d$ as small as 0.1 and 0.12, the fence cannot retain any sphere. Therefore, $n_c$ stays at zero and $\bar{z}_c$ at its initial value.

Finally, we systematically varied the spheres' stiffness, as presented in Fig. \ref{Fig_Stiffness}. The stiffness was written as a dimensionless variable as $k_n/Pd$, where $P$ corresponds to the hydrostatic pressure at the bottom of the system (see section IV.A in \cite{dacruz2005rheophysics}). The results show comparable mean values and dispersion in a wide range of $k_n/(Pd)$ (from 10 to 3000). This is particularly noticeable in $\bar{z}_c$ (Fig.~\ref{Fig_Stiffness}(a)). For $n_c$ (Fig. \ref{Fig_Stiffness}(b)), one may argue a decreasing tendency starting from $k_n/(Pd)$ = 1000. However, the error bars show that until $k_n/(Pd)$ = 3000, we are still close to the constant value obtained in our model. 
For even larger $k_n/(Pd)$ values, the strong rigidity of the spheres makes the system less stable and more sensitive to perturbations, where the collapse happens rapidly. However, this infinitely rigid case is less justified when compared to realistic situations.

In summary, our results vary slightly regarding friction, fence height, and spheres' stiffness. Therefore, the scenario presented in this article is expected to qualitatively reproduce the phenomena observed in nature, particularly in the example from the market shown in Fig.~\ref{fig:wide}.

%%%%%%%%%%%%%%% Conclusions
\section{\label{sec:Conclusions} Conclusions}
%\paragraph{Conclusions. \label{sec:Conclusions}}
We conducted discrete element simulations to study avalanches triggered by vacancies in a granular medium formed by a monolayer of crystallized spheres on a tilted plane. By varying the inclination angle of the plane and the size of the system, a phase diagram was built where three zones are distinguished: \textit{no avalanche}, \textit{avalanches triggered by extractions}, and \textit{instant avalanches}. We observed that, as particles are extracted, the average coordination number ($\bar{z}$) decreases, weakening the system's structure.
On the other hand, we showed that in most of the \textit{avalanche triggered by extractions} zone, the critical extraction number $n_c$ can be predicted by a simple statistical model of the average coordination number. The prediction is made by adding the critical coordination number obtained from DEM simulations to our evolution model.  Only a few cases at transition zones are out of the model's predictions, thus requiring a more specific study.\\
Three critical dimensionless constants were discovered in the present work. The first one is the mean critical coordination number $\bar{z}_c=4.27$, which represents a limit of network connection under which the system collapses. The second one is a critical void ratio $n_c/N_0=0.10$, which gives us a critical packing fraction $\phi_c=0.82$, very close to the limit of the flowing region in simple shear. 

Our numerical simulations and the proposed model shed some light on a problem that is difficult to tackle experimentally. We should investigate if our results could be extended to other situations, such as non-crystallized poly-disperse systems or 3D configurations.
Finally, we want to emphasize the value of considering everyday phenomena to inspire new pathways to learn about complex problems like avalanches.\\

\begin{acknowledgments}
%\paragraph{Acknowledgments.}
The authors thank Nader Droguett, Gabriel Maureira, and Thomas Olivares for performing preliminary experiments that inspired this study. The authors also thank Agencia Nacional de Investigación y Desarrollo (ANID-Chile) for financially supporting this research through grants ING2030, 16ENI2-71940 (E.R.), as well as Fondecyt Grants: 11230970 (E.R.), 11190900 (V.S.), 11200464 (G.C.), and 11191106 (P.G.).
\end{acknowledgments}

%\nocite{*}

%\bibliography{MonolayerAvalanches}% Produces the bibliography via BibTeX.

%------------------------------------
\bibliographystyle{apsrev4-1.bst}

\end{document}